\documentclass[times,authoryear]{elsarticle}

\usepackage{jasr}
\usepackage{framed,multirow}

\usepackage{amssymb}
\usepackage{latexsym}

\usepackage{tabularx}
\usepackage{graphicx}
\usepackage{amsmath}
\usepackage{multirow}
\usepackage{subcaption}
\usepackage{float}

\usepackage[switch]{lineno}

\usepackage{url}
\usepackage{xcolor}
\definecolor{newcolor}{rgb}{.8,.349,.1}

\usepackage[citebordercolor=white]{hyperref}

\journal{Advances in Space Research}

\begin{document}

\verso{Ziyadat Hassan \textit{etal}}

\begin{frontmatter}

\title{Investigating Ionospheric TEC Variations in Solar and Geomagnetic Influences Across Solar Activity Phases}

\author[1,2]{Ziyadat \snm{Hassan}}
\ead{ziyadathassan@gmail.com}
\author[1]{Zamri \snm{Zainal Abidin}\corref{cor1}}
\cortext[cor1]{Corresponding author: 
  Tel.: +0-000-000-0000;  
  fax: +0-000-000-0000;}
\ead{zzaa@um.edu.my}
\author[1]{Affan Adly \snm{Nazri}}
\author[1]{Nursyazela Badrina \snm{Baharin}}

\affiliation[1]{organization={Radio Cosmology Research Laboratory, Center for Astronomy and Astrophysics Research},
                addressline={Department of Physics, Faculty of Science, Universiti Malaya},
                city={Kuala Lumpur},
                postcode={50603},
                country={Malaysia}}
\affiliation[2]{organization={Department of Physics},
                addressline={School of Science, Adamu Augie College of Education},
                city={Argungu, Kebbi State},
                postcode={P.m.b. 1012},
                country={Nigeria}}

\received{xx xxx 2025}
\finalform{xx xxx 2025}
\accepted{xx xxx 2025}
\availableonline{xx xxx 2025}
\communicated{}

\begin{abstract}
This study examines the variability of ionospheric total electron content (VTEC) in response to solar and geomagnetic drivers across solar cycles 23 to 25. While the dominant effect of solar radiation on VTEC is well-known, a comprehensive understanding of how these relationships and their time-lags vary across distinct solar cycle phases and across cycles of differing intensity has been lacking. Using global VTEC data from the Chinese Academy of Sciences Global Ionospheric Maps (CASG) and solar-geophysical indices from NASA’s OMNI dataset spanning from 1998 to 2025, this study bridges that gap by quantifying correlation strengths and time-lag relationships between VTEC and parameters such as $F_{10.7}$ solar flux, R sunspot number, Kp, Ap, and Dst indices, and solar wind properties. Results show that solar proxies, particularly $F_{10.7}$ and R sunspot number, exhibit the strongest, most consistent correlations with VTEC, especially during the ascending and descending phases of the solar cycle, with a characteristic $\sim2$-day lag attributed to thermospheric oxygen dynamics and ionospheric recombination processes. In contrast, geomagnetic indices exhibit weaker and phase-dependent correlations, while direct correlations between solar wind parameters and global VTEC are weak, as their influence is primarily mediated by geomagnetic activity and exhibits strong regional and temporal heterogeneity. Phase-resolved analyses further reveal that geomagnetic activity plays a more prominent role during transitional phases, while maximum and minimum periods are dominated by EUV variability and non-solar drivers, respectively. These findings highlight the necessity of incorporating solar phase and time-lag dependencies in ionospheric modelling and forecasting efforts.
\end{abstract}


\end{frontmatter}


\section{Introduction}\label{sec:intro}

The ionosphere, a dynamic region of Earth's upper atmosphere ($\sim 60 - 1000~\mathrm{km}$ altitude) plays a critical role in trans-ionospheric radio signal propagation. Its behaviour, characterised by spatial and temporal variability in free electron density \citep{Dabas2000}, directly impacts satellite communications, global navigation satellite systems (GNSS), position accuracy, and radio astronomy \citep[e.g.][]{Nazri2025}. Total electron content (TEC), defined as the integrated electron count along a ray path (units: TECU, $1~\mathrm{TECU} = 10^{16}~\mathrm{electrons}~\mathrm{m}^{-2}$), usually between a ground receiver and satellite \citep{Bust2008}, serves as a key parameter for quantifying ionospheric variability \citep{Ciraolo2002}. TEC measurements are indispensable for correcting ionospheric delays in GNSS applications \citep{Wang2013}, modelling space weather impacts, and predicting ionospheric disturbances such as equatorial plasma bubbles \citep{Khamdan2021} and mid-latitude anomalies.

Ionospheric TEC fluctuations are predominantly driven by solar activity via extreme ultraviolet (EUV) radiation and solar wind \citep{Parwani2021} while also being influenced by geomagnetic activity that is triggered by solar wind-magnetosphere interactions \citep{Latovika1996}. Solar EUV radiation ionises neutral atoms, increasing electron production \citep{Liu2011}, while geomagnetic storms redistribute plasma through electrodynamic processes \citep{Immel2013}. The $F_{10.7}$ index \citep{Qian2025} and sunspot number \citep[R;][]{Zossi2025} are well-established proxies for solar EUV output and correlate strongly with TEC over solar cycles \citep{Woldemariam2024}. Conversely, geomagnetic indices (e.g. Kp, Dst) modulate TEC through storm-induced thermospheric heating \citep{Yamazaki2024} and composition changes \citep{Sahai2011}, though correlations are complex \citep{Klimenko2017} and regionally heterogeneous \citep{Mukhtarov2025}.

Despite advances, critical research gaps persist. When discussing phase-dependent responses, prior studies lack consensus on how TEC correlations with solar/geomagnetic drivers vary across solar cycle phases (ascending, maximum, descending, minimum). As an example, while it has been established that solar EUV flux strong ionises the ionosphere \citep{Reid1976}, frequent coronal mass ejections (CMEs) and solar flares can cause conditions with high Kp and negative Dst that disrupt the EUV-TEC relationship through thermospheric heating as reported by \cite{Oljira2023} in Tromso and plasma redistribution even beyond the ionosphere as shown by \cite{Liu2021}, resulting in weaker TEC-EUV correlation despite occurring during solar maxima. \cite{Jee2014} on the other hand, showed how EUV irradiance is consistently low during solar minima, but \cite{Borries2024} reported deviations from the norm during solar quiet years of 2005 -- 2006 through a tenable Joule heating mechanism as derived by \cite{Palmroth2004}. And while EUV dominates during solar maxima, the role of high-speed solar wind streams from coronal holes during declining phases remains under-explored as conceded by \cite{Verbanac2011}.

Regarding time-lag dynamics, the ionosphere response time to solar/geomagnetic forcing is inconsistently quantified, even with attempts at quantifying such sluggishness as what was done by \cite{Chakraborty2021}. While $F_{10.7}$ exhibits near-immediate effects, lags for sunspot-driven and geomagnetic responses are poorly characterised across cycles \citep{Liu2006b}. Most studies reviewed also focus on single solar cycles, and comparative analyses spanning multiple cycles are scarce, as was categorically stated by \cite{Riley2023}. We build up a study to bridge these gaps by quantifying phase-resolved correlations between TEC and solar/geomagnetic parameters across solar cycles 23 -- 25 and by evaluating time-lag dependencies to resolve ionospheric response timescales. We organise the paper as follows: Section 2 describes the TEC, solar, and geomagnetic datasets used, as well as the correlation and time-lag methodology. The results and associated physical phenomena are presented and discussed in Section 3. 

\section{Methodology}\label{sec:method}

\subsection{CAS Global Ionospheric Map Data}\label{sec:method:casg}

To characterise the ionosphere, we used global TEC maps produced by the Chinese Academy of Sciences \citep[CAS; hereafter referred to as CASG maps;][]{Li2021}. These maps are generated from GNSS data, and extracts ionospheric variables by utilizing a geometry-free combination of dual-frequency pseudorange and carrier phase observations \citep{Afraimovich2013}. The data were further refined using the Carrier-to-Code Levelling (CCL) technique \citep{Mannucci1998}. Vertical TEC (VTEC) is then modelled using the Spherical Harmonic Plus Thin Shell (SHPTS) approach, which combines global-scale spherical harmonic (SH) expansions \citep{Schaer1999} with local-scale modified generalized trigonometric series (MGTS) \citep{Yuan2019}. The resulting VTEC values are gridded based on the International GNSS Service (IGS) convention using the Differential Areas for Differential Stations (DASD) method \citep{Yuan2002}, with temporal resolution up to 30 minutes. The final CASG maps are expressed in TEC units (TECU), where 1 TECU corresponds to $10^{16}$ electrons per square meter. To ensure continuity across day boundaries, the final CASG VTEC products incorporate an additional 4 hours of observations before and after each day.

\subsection{NASA OMNI Solar and Geomagnetic Parameters}\label{sec:method:omni}

To characterise solar and geomagnetic conditions relevant to our study, we employed the OMNI 2 dataset curated by the National Aeronautics and Space Administration \citep[NASA;][]{King2005}. OMNI compiles near-Earth solar wind and geomagnetic data from over 20 spacecrafts, in conjunction with measurements from various ground-based observatories. From this dataset, we extracted the following parameters: 
\begin{enumerate}
    \item Solar wind proton density, plasma speed, and flow pressure from NASA's Wind, ACE, IMP, and Geotail satellites;
    \item R Sunspot number from the Belgium Solar Influences Data Analysis Center \citep{SILSO};
    \item 10.7~cm solar flux ($F_{10.7}$) from the Canadian Space Agency\footnote{\url{https://www.spaceweather.gc.ca/forecast-prevision/solar-solaire/solarflux/sx-en.php}}.
    \item Planetary K- and A-index (Kp and Ap indices) from the German Research Centre for Geosciences\footnote{\url{https://kp.gfz.de/en/data}}; and
    \item Disturbance storm time (Dst) index from the World Data Center for Geomagnetism\footnote{\url{https://wdc.kugi.kyoto-u.ac.jp/}}.
\end{enumerate}

Solar wind proton density, plasma speed, flow pressure, R sunspot number, and $F_{10.7}$ serve as indicators of solar activity. The first three parameters describe the properties of the solar wind i.e. the plasma stream emitted from coronal holes in the Sun. The R sunspot number provides a direct count of visible sunspots on the solar surface, and is indicative of magnetic activity on the solar surface. $F_{10.7}$ index represents the solar radio flux at a wavelength of 10.7~cm, emitted from both the upper chromosphere and lower corona, and is widely used as a proxy for solar activity due to its reliability and insensitivity to terrestrial weather conditions. The Kp, Ap, and Dst indices, in contrast, quantify geomagnetic activity. The Kp and Ap indices measure global disturbances in the horizontal component of Earth's magnetic field in quasi-logarithmic and linear scales respectively, while the Dst index captures the intensity of the ring current through the hourly average perturbation of the geomagnetic field at low latitudes. 

\subsection{Data Analysis}\label{sec:analysis}

\subsubsection{VTEC and OMNI Data Averaging}\label{sec:analysis:tec}

This study made use of VTEC data from the CASG maps and solar/geomagnetic indices from the OMNI dataset, covering the period from 1 January 1998 to 31 January 2025. To minimize the influence of diurnal variation and spatial dependence in ionospheric behaviour, CASG maps were averaged into daily VTEC values. These daily VTEC measurements, alongside the daily OMNI data, form the basis for subsequent time lag analyses (see Section~\ref{sec:analysis:lag}). For other direct comparisons and statistical analyses, the data were further aggregated on a monthly basis. We note that this approach, while optimal for identifying long-term, global-scale trends and phase-dependent responses across solar cycles, necessarily masks finer-scale latitudinal, longitudinal, and diurnal dependencies in ionospheric behaviour. Future studies employing local time and latitudinally resolved analyses would be valuable to explore these important regional and diurnal effects.

We define solar activity phases according to the sunspot cycle phase function proposed by \cite{Maliniemi2014}, with phase delineations adapted following the approach described by \cite{Getachew2017}. Each solar cycle is divided into four phases centred at phase values of $0$, $\pi/2$, $\pi$, and $3\pi/2$, representing minimum, ascending, maximum, and descending phases respectively, with each phase spanning a width of $\pi/2$. These phase boundaries are determined by linearly interpolating between solar minima and maxima based on the 13-month smoothed sunspot number data \citep{SILSO}. The corresponding time intervals for each phase within solar cycles 23 – 25 are presented in Table~\ref{tab:phases}. Note that the maximum of solar cycle 25 is based on the current tentative peak (October 2024). On the other hand, the minimum for solar cycle 26 is interpolated based on predictions for the cycle's peak using machine learning algorithms trained based on various solar parameter data \citep{Liu2023, Kalkan2023, Luo2024, Cao2024, Rodrguez2024a, Zeng2025}.

\begin{table}
    \centering
    \caption{Range of months of solar cycle phases in solar cycles 23 to 25.}
    \label{tab:phases}
    \def\arraystretch{1.2}
    \begin{tabular}{cccc}
        \hline
        Cycle & Phase & Start & End \\
        \hline
        \multirow{3}{*}{23} & Ascending & December 1997 & July 2000 \\
        & Maximum & August 2000 & August 2003 \\
        & Descending & September 2003 & February 2007 \\
        \hline
        \multirow{4}{*}{24} & Minimum & March 2007 & April 2010 \\
        & Ascending & May 2010 & November 2012 \\
        & Maximum & December 2012 & August 2015 \\
        & Descending & September 2015 & July 2018 \\
        \hline
        \multirow{3}{*}{25} & Minimum & August 2018 & February 2021 \\
        & Ascending & March 2021 & July 2023 \\
        & Maximum & August 2023 & January 2026 \\
        \hline
    \end{tabular}
    
\end{table}

\subsubsection{Quantifying Correlation}\label{sec:analysis:correlation}

To evaluate the relationship between VTEC and selected solar and geomagnetic parameters, we use Pearson's correlation coefficient, which is given by 
\begin{equation}
    r = \frac{\sum_{i=1}^n (x_i-\bar{x})(y_i-\bar{y})}{\sqrt{\sum_{i=1}^n (x_i-\bar{x})^2} \sqrt{\sum_{i=1}^n (y_i-\bar{y})^2}}
\end{equation}
where $x_i$ denotes the VTEC values, $y_i$ represents the corresponding solar or geomagnetic parameter values, and $n$ is the sample size. This coefficient is computed for the full dataset and separately for data subsets corresponding to each solar activity phase (see Section~\ref{sec:analysis:tec}). To evaluate the statistical significance of the observed correlation, we also examined the corresponding p-values assuming normally distributed populations for both parameters. Any relation with $p < 0.05$ was considered statistically significant (and vice versa), and here we report the p-value in their base-10 logarithms (generally shown in brackets), which corresponds to $\log_{10}p < -1.3$.

Note that Pearson's correlation coefficient mainly measures the degree of linear correlation in the data. Alternatively, one could employ the Spearman's rank correlation coefficient which is a non-parametric measure of rank correlation, and only assesses the monotonicity of the relationship which could reveal any non-linear connection between the two variables. The Spearman's correlation coefficient is given by
\begin{equation}
    \rho = \frac{\mathrm{cov}\left[R[X], R[Y]\right]}{\sigma_{R[X]} \sigma_{R[Y]}} = 1 - \frac{6\sum_{i=0}^n (R[X_i] - R[Y_i])^2}{n(n^2 - 1)}
\end{equation}
where $R[X_i]$ and $R[Y_i]$ are the ranks of respective values. Similarly, Kendall's rank correlation coefficient measures ordinal association by the method of concordance-discordance fractions, which is explicitly expressed as
\begin{equation}
    \tau = \frac{2}{n(n-1)} \sum_{i < j} \mathrm{sgn}(x_i - x_j)~\mathrm{sgn}(y_i - y_j)
\end{equation}
where $\mathrm{sgn}$ is the sign function. However, across both full and phase-filtered datasets, we consistently found that $\lvert r\rvert > \lvert \rho \rvert > \lvert \tau \rvert$, indicating that the correlations (if present) are primarily linear. Therefore, we prioritized Pearson's $r$ in our analysis.  

\subsubsection{Time Lag Calculations}\label{sec:analysis:lag}

To explore potential lagged responses of the ionosphere to variations in solar and geomagnetic conditions, we applied a time-shift analysis using the daily averaged VTEC and OMNI datasets. In this analysis, the VTEC time series was systematically shifted with respect to the OMNI time series by integer lag values from $-30$ to $+30$ days (i.e. up to 30 days earlier or later). For each lag value, the shifted VTEC data and the OMNI data were aligned over their overlapping time interval, and the correlation coefficients were recalculated following the procedure described in Section~\ref{sec:analysis:correlation}. The resulting correlation–lag relationship enabled us to identify whether the strongest coupling occurs with a delay (positive lag), in advance (negative lag), or without any lags.

\section{Results and Discussion}\label{sec:results}

\subsection{Ionospheric, Solar, and Geomagnetic Parameters Temporal Variations}\label{sec:results:temporal}

\begin{figure}
    \centering
    \includegraphics[width=0.95\linewidth]{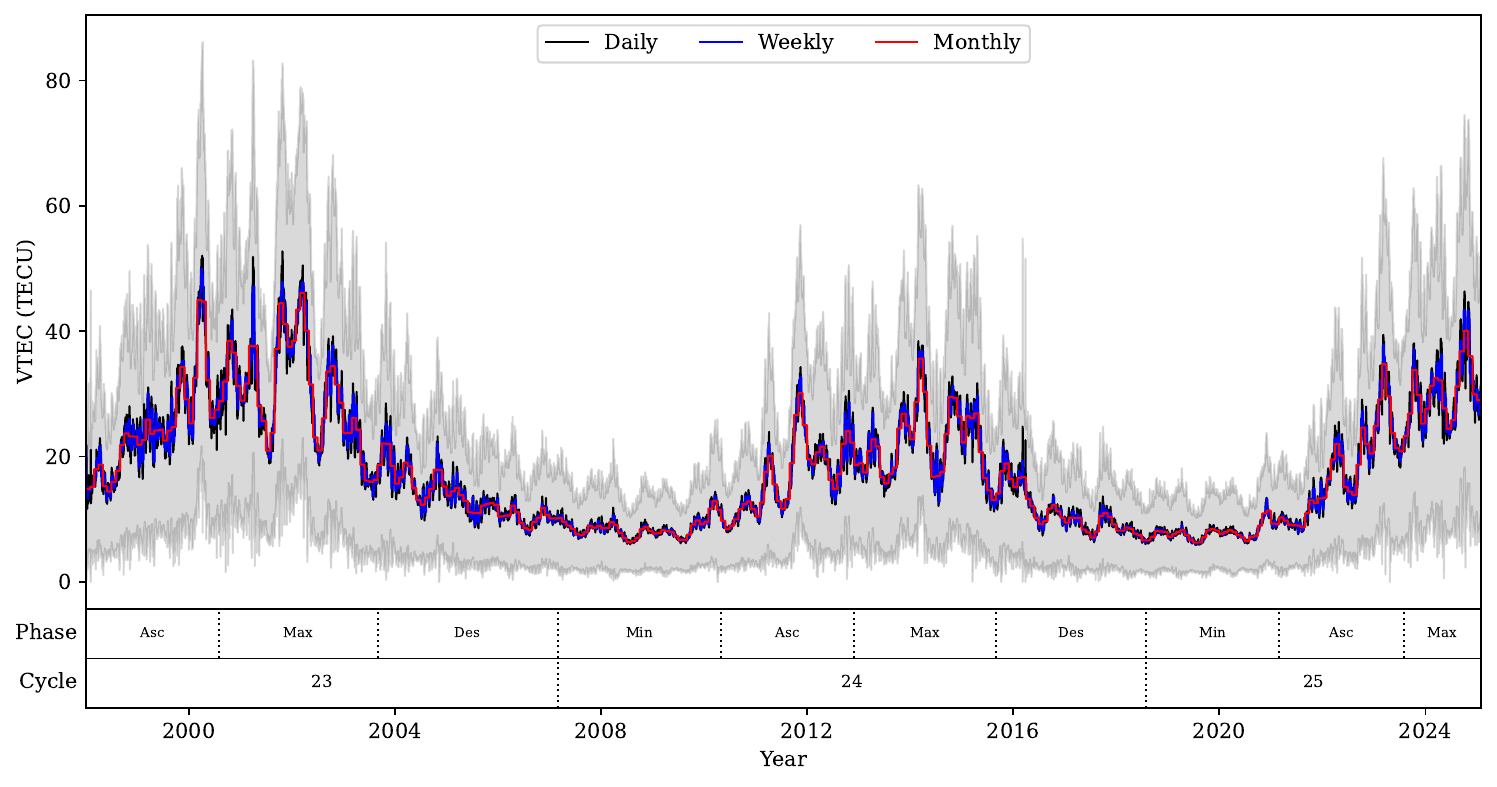}
    \caption{Long-term variation of VTEC across multiple solar cycles (1998 -- 2025). Black, blue, and red lines represent the daily, weekly, and (calendar) monthly averaged VTEC values, with the grey region denoting the VTEC standard deviation on a specific day. The lower two frames show the traditional solar phases (Asc: Ascending, Max: Maximum, Des: Descending, Min: Minimum) and solar cycles.}
    \label{fig:vtec}
\end{figure}

Shown in Figure \ref{fig:vtec} is the long-term variation of VTEC values from 1998 to 2025. Naturally, the VTEC levels are higher during solar maximum periods, and lower during solar minima. These trends are in line with well-established solar cycle effects, where higher electron absorptions are the result of enhanced ionospheric ionisation caused by increased solar irradiation and geomagnetic activity \citep{Kundu2021}. 

Interestingly, solar cycle 23 exhibits the largest VTEC scales, followed by lower values in cycle 24, with a gradual increase in cycle 25. This tallies with established solar activity, indicating a rebound in the ionospheric response to solar forcing. This also may indicate the presence of the Centennial Gleissberg Cycle \citep[CGC;][]{Gleissberg1939} i.e. a $\sim 100$ year cycle superimposed on the general 11-year solar cycle and resulting in a significantly weaker solar maxima every seven to eight years. In this case, the CGC minimum appears to have occurred in cycle 24, which is consistent with other analyses \citep{Adams2025}.

\begin{figure}
    \centering
    \includegraphics[width=0.75\linewidth]{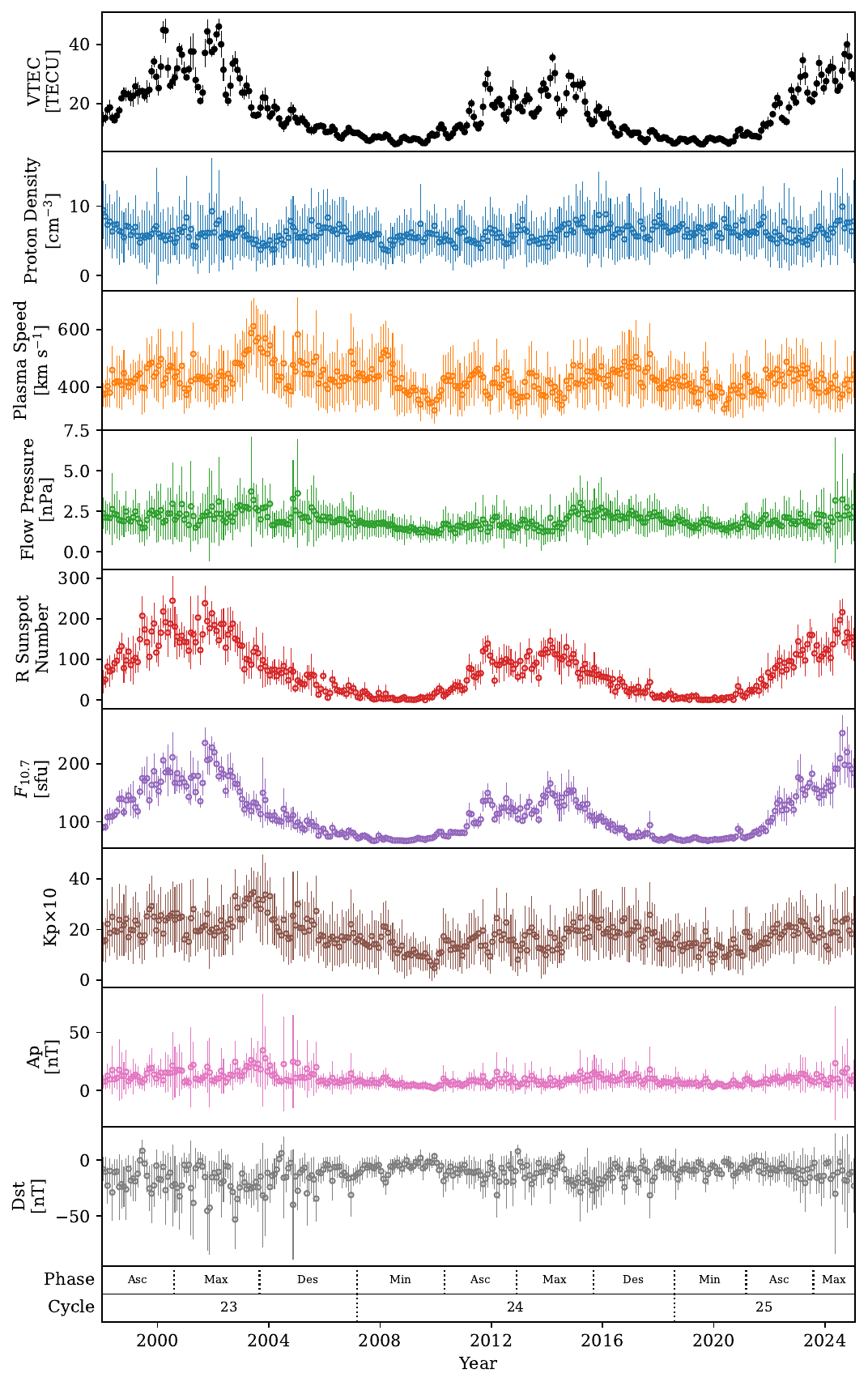}
    \caption{Long-term variation of the solar and geomagnetic parameters in comparison with VTEC. The points in each plot show the monthly averaged data with error bars signifying the standard deviation of the respective parameters. The lowest two frames depict the traditional solar phases (Asc: Ascending, Max: Maximum, Des: Descending, Min: Minimum) and solar cycles.}
    \label{fig:vtec_omni}
\end{figure}

Figure \ref{fig:vtec_omni} displays the long-term variation of solar and geomagnetic parameters in comparison with VTEC over the three solar cycles. The sunspot number (R) and $F_{10.7}$ solar radio flux demonstrate similarities to VTEC, with all three parameters exhibiting pronounced peaks and troughs in solar maxima and minima of each solar cycle respectively. The exact correlations and reasoning are discussed further in Section \ref{sec:results:corr}. 

In contrast, other solar and geomagnetic parameters do not exhibit clear resemblance with the evolution of VTEC. One exception is the Dst index, which shows significant negative excursions, particularly during solar maxima phases, which shows weak negative correlation with VTEC (See Section \ref{sec:results:corr}). Another interesting feature seen in the data is the peak in solar plasma speed, flow pressure, $F_{10.7}$, Kp index, and Ap index at around 2004, but not in VTEC levels. This time frame corresponds to the ``2003 Halloween Storm'' where a series of solar flares and CMEs from mid-October to early November 2003 resulted in intense geomagnetic storms with effects lingering into early 2004 \citep{Hu2023}. However, the peaking of VTEC due to the storm (strongest on 28 - 29 October 2003) is not seen in our monthly averaged data due to the quick response of the ionosphere which recovered within a day \citep{Tsurutani2006a}. These relationships highlight the complex interplay between solar activity and ionospheric behaviour.

\subsection{Correlations and Non-correlations}\label{sec:results:corr}

\begin{figure}
    \centering
    \begin{tabular}{cc}
        \includegraphics[width=0.37\textwidth]{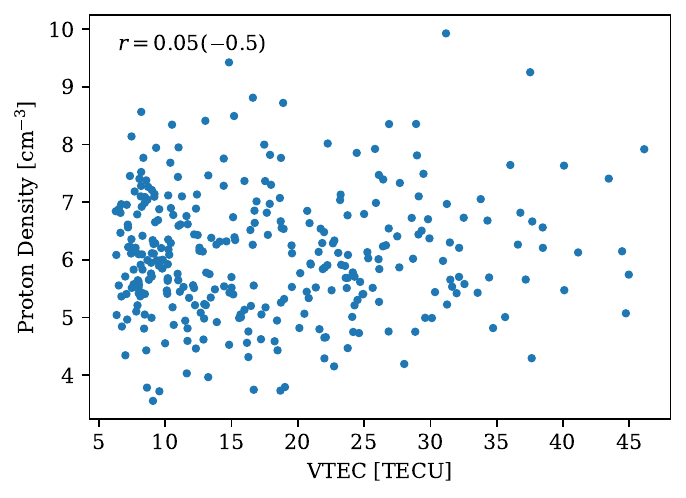} & \includegraphics[width=0.37\textwidth]{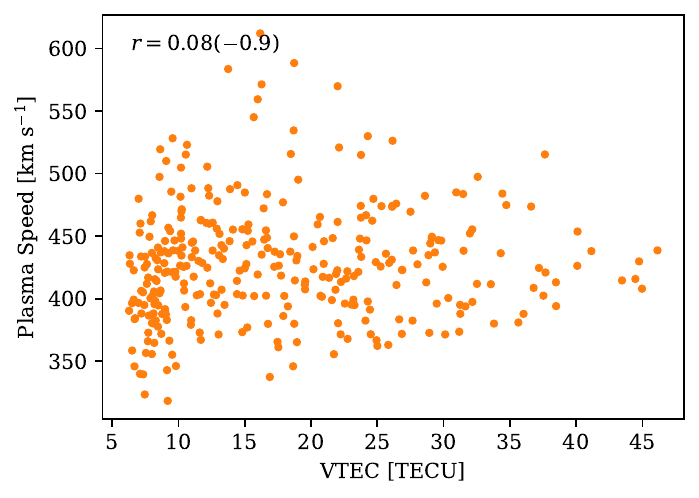} \\
        \includegraphics[width=0.37\textwidth]{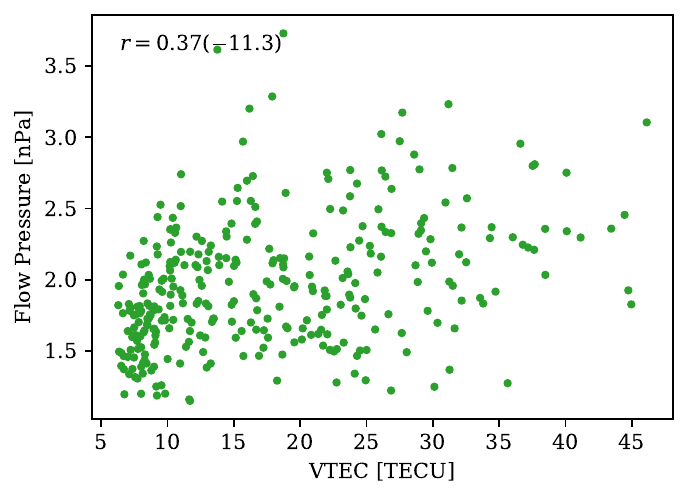} & \includegraphics[width=0.37\textwidth]{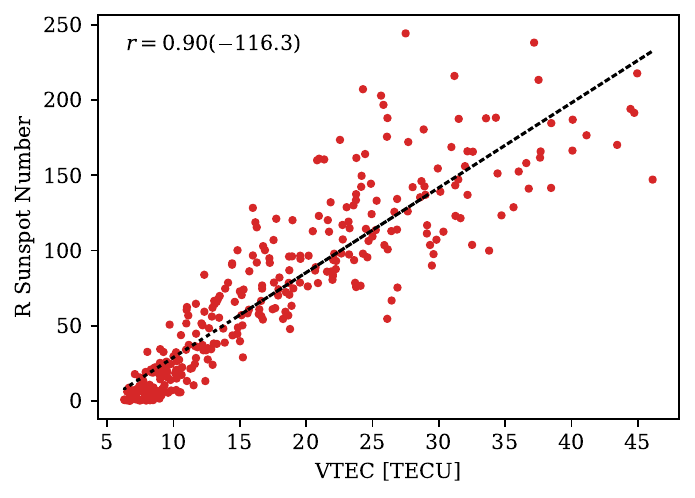} \\
        \includegraphics[width=0.37\textwidth]{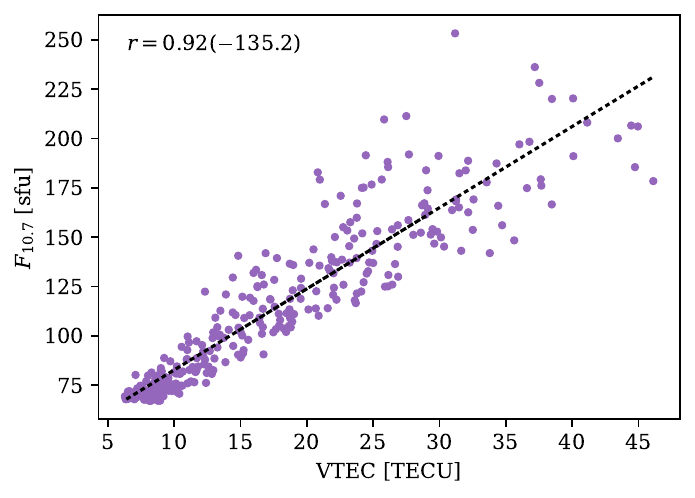} & \includegraphics[width=0.37\textwidth]{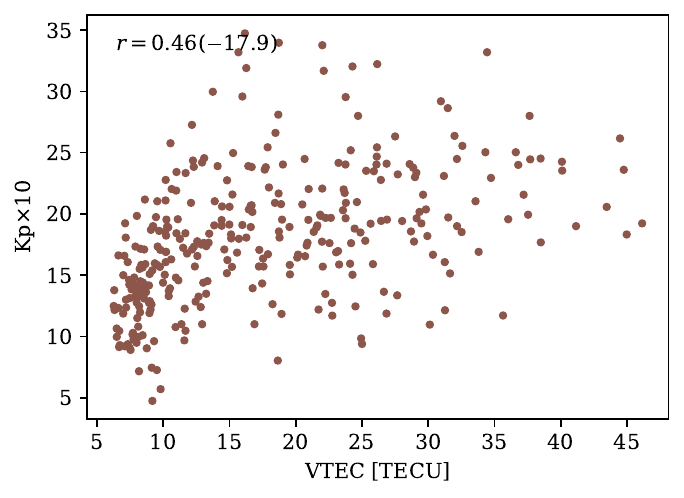} \\
        \includegraphics[width=0.37\textwidth]{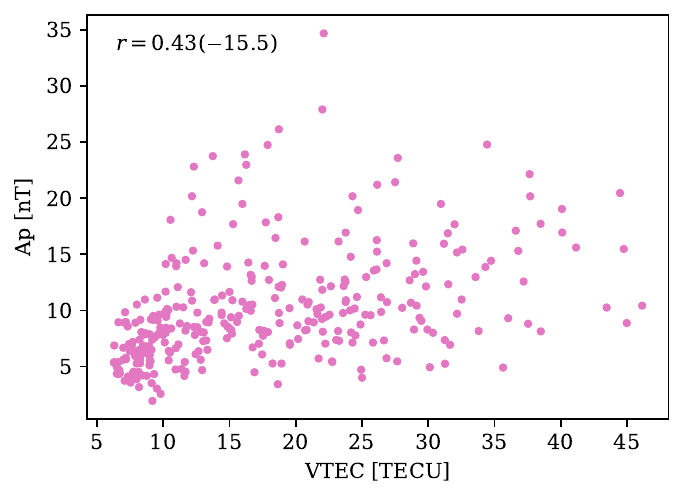} & \includegraphics[width=0.37\textwidth]{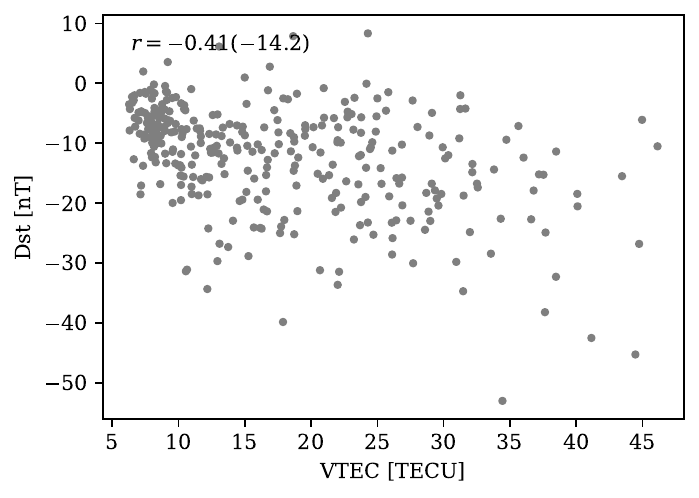}
    \end{tabular}
    \caption{Comparison of VTEC with solar and geomagnetic parameters. The Pearson's correlation coefficient for each relationship is shown in the top left of each plot. The values shown in the brackets correspond to the base-10 logarithm of the respective p-values. For relations with a correlation coefficient of $r > 0.5$, a dashed line depicts a linear regression fitted to said data.}
    \label{fig:vtec_corr}
\end{figure}

Figure \ref{fig:vtec_corr} shows the comparison between VTEC values with solar and geomagnetic parameters over the entire data range. The R sunspot number and $F_{10.7}$ solar flux exhibit strong correlations with VTEC, with Pearson's correlation coefficients of 0.90(-116.3) and 0.92(-135.2) respectively. Moderate correlations exist for geomagnetic parameters i.e. coefficients of 0.46(-17.9) for Kp index, 0.43(-15.5) for Ap index, and -0.41(-14.2) for Dst index, while the solar wind parameters i.e. proton density, plasma speed, and flow pressure exhibit weak to no correlation in comparison with VTEC, possessing coefficients of 0.05(-0.5), 0.08(-0.9), and 0.37(-11.3) respectively. 

The R sunspot number and the $F_{10.7}$ solar radio flux both exhibit strong correlations with VTEC, highlighting the dominant role of solar activity in modulating ionospheric electron density. These parameters consistently peak during solar maxima, reflecting periods of enhanced solar ultraviolet (UV) and extreme ultraviolet (EUV) radiation that drive increased ionisation in the ionosphere, thereby elevating VTEC values \citep{Liu2006a, Chen2011}. Quantitatively, correlations between VTEC and the R sunspot number as well as $F_{10.7}$ flux affirm this strong coupling, particularly under geomagnetically quiet conditions \citep{Liu2006b, Lissa2021}. While the $F_{10.7}$ index serves as a high-resolution proxy for solar EUV output, enabling short- and long-term VTEC predictions, the sunspot number provides broader context regarding the solar cycle phase. These complementary indices have been validated by empirical and modelling studies as reliable predictors of ionospheric variability \citep{Kassa2017}.

The moderate correlation between VTEC and the geomagnetic parameters indicates that while geomagnetic activity does influence the atmosphere, its impact on VTEC is less direct and less consistent than that of solar effects. For Kp and Ap, the weak relationship arises due to the temporal mismatch i.e. Kp and Ap measures geomagnetic disturbances over 3-hour intervals, whereas VTEC is more responsive to continuous solar radiation inputs. Nevertheless, during geomagnetic activity, enhanced magnetosphere-ionosphere coupling can strengthen the equatorial fountain effect through processes such as Joule heating, particle precipitation, and prompt-penetration electric fields (PPEFs), thereby increasing plasma uplift and electron density at low latitudes. This results in localized TEC enhancement, commonly known as the ``positive storm effect'' \citep{Bhaskar2013, Fejer2024}. The complex, non-linear, and region-dependent ionospheric response to geomagnetic activity further displaces the correlation \citep{Astafyeva2015}. 

In contrast, Dst often exhibits negative excursions during solar maxima as a result of intense geomagnetic storms driven by CMEs \citep{Bhaskar2013}. The intensification of the ring current during such storms enhances thermospheric heating and alters composition (e.g. decreasing the O/N$_2$ ratio), which increases recombination and depletes ionospheric plasma, and this is commonly referred to as the ``negative storm effect'' \citep{Immel2013, Zhai2023}. These effects are particularly evident at mid-latitudes during the storm recovery phase \citep{Astafyeva2015}. Overall, the TEC response to geomagnetic forcing is highly dependent on both latitude and storm phase, with positive responses dominating in equatorial regions during main-phase disturbances and negative responses prevailing in mid-latitudes during recovery \citep{Tsurutani2006b, Verbanac2011}.

As for solar wind parameters, none of them show a strong, direct global correlation with VTEC, except for solar wind dynamic pressure which exhibits a weak relationship. This outcome is consistent with the understanding that the solar wind's primary influence on the ionosphere is indirect, mediated by its triggering of geomagnetic activity (e.g., through indices like Kp, Ap, and Dst), rather than through a direct, globally uniform causal link. During geomagnetically disturbed periods, however, studies have shown that VTEC can intensify, often initiated by solar wind structures such as corotating interaction regions (CIRs) and high-speed streams (HSSs) through enhanced magnetospheric compression and penetration electric fields \citep[e.g.][]{Verkhoglyadova2011, Mishra2020}. This response, nevertheless, varies significantly with latitude, local time, season, and solar cycle phase \citep{FullerRowell1997}. For instance, at high latitudes, VTEC levels tend to increase under strong solar wind energy input during winter nights, but show an inverse reaction during summer days due to contrasting thermospheric and electrodynamic conditions \citep{Borries2024}. At low and equatorial latitudes, VTEC is modulated by solar wind variations, though the response is typically weaker or delayed relative to higher latitudes \citep{Mishra2020, Fejer2024}. With all matters considered, to average over relatively long timescales and over the entire Earth's surface would inevitably disrupt the diverse responses between solar wind properties and VTEC, finally obscuring any correlation that may have existed.

\subsection{Solar Activity Phase Dependence}\label{sec:results:phase}

Table \ref{tab:corr-phase} demonstrates the dynamic nature of the correlation between VTEC with the solar and geomagnetic parameters throughout the different solar activity phases, generally with high statistical significance. A more persistent and direct ionospheric reaction during ascending and descending phases are seen for R sunspot number and $F_{10.7}$, mainly due to moderate and gradual nature of the change in solar UV flux. Conversely, in solar maxima phases, disturbances such as increased geomagnetic activities and solar activities (e.g. CMEs and solar flares) result in major disruptions to the ionosphere which complicate the direct impact of solar parameters on VTEC, hence the slightly weaker correlation \citep{Afraimovich2013, Tsurutani2006b}. Similarly, solar minima phases causes the ionosphere to experience lower ionisation as a result of the low solar radiation levels. This causes VTEC to be more susceptible to non-solar factors e.g. atmospheric waves and neutral dynamics \citep{Afraimovich2008}. Recent studies further show that planetary waves, atmospheric tides (both solar and lunar), and coupling from the lower atmosphere can significantly modulate TEC during weak solar activity, with tidal and lunar components becoming particularly discernable during deep minima or sudden stratospheric warming events \citep{Zhai2023, Hocke2024, Ma2025}.

\begin{table}
    \centering
    \caption{Correlation values for each solar activity phase, for each solar and geomagnetic parameter against VTEC. The values shown in the brackets correspond to the base-10 logarithm of the respective p-values.}
    \label{tab:corr-phase}
    \footnotesize
    \def\arraystretch{1.2}
    \begin{tabular}{cccccccccc}
        \hline
        Phase & Cycle & Proton Density & Plasma Speed & Flow Pressure & R Sunspot No. & $F_{10.7}$ & Kp Index & Ap Index & Dst Index \\
        \hline
        All data & 23 - 25 & 0.05(-0.5) & 0.08(-0.9) & 0.37(-11.3) & 0.90(-116.3) & 0.92(-135.2) & 0.46(-17.9) & 0.43(-15.5) & -0.41(-14.2) \\\hline
        \multirow{3}{*}{Ascending} & 23 & -0.54(-2.8) & 0.37(-1.4) & -0.04(-0.1) & 0.73(-5.4) & 0.79(-6.8) & 0.32(-1.1) & 0.22(-0.6) & -0.10(-0.2) \\
         & 24 & 0.25(-0.8) & -0.48(-2.2) & 0.03(-0.1) & 0.86(-9.3) & 0.86(-9.1) & 0.05(-0.1) & 0.16(-0.4) & -0.25(-0.7) \\
         & 25 & -0.50(-2.2) & 0.55(-2.7) & 0.18(-0.4) & 0.79(-6.5) & 0.85(-8.2) & 0.76(-5.9) & 0.78(-6.2) & -0.60(-3.3) \\\hline
        \multirow{3}{*}{Maximum} & 23 & 0.43(-2.1) & -0.56(-3.4) & -0.10(-0.2) & 0.58(-3.8) & 0.75(-6.9) & -0.37(-1.6) & -0.16(-0.5) & -0.31(-1.2) \\
         & 24 & -0.02(-0.0) & 0.04(-0.1) & 0.13(-0.3) & 0.50(-2.5) & 0.71(-5.4) & 0.01(-0.0) & -0.04(-0.1) & -0.16(-0.4) \\
         & 25 & 0.16(-0.3) & 0.04(-0.1) & 0.17(-0.3) & 0.03(-0.0) & 0.30(-0.6) & 0.35(-0.8) & 0.29(-0.6) & -0.29(-0.6) \\\hline
        \multirow{2}{*}{Descending} & 23 & -0.51(-3.3) & 0.48(-2.9) & 0.26(-1.0) & 0.75(-8.1) & 0.83(-11.0) & 0.68(-6.2) & 0.62(-4.9) & -0.29(-1.2) \\
         & 24 & 0.28(-1.0) & 0.13(-0.3) & 0.66(-4.8) & 0.82(-8.6) & 0.84(-9.6) & 0.61(-4.0) & 0.55(-3.2) & -0.62(-4.1) \\\hline
        \multirow{2}{*}{Minimum} & 24 & -0.09(-0.2) & 0.01(-0.0) & 0.05(-0.1) & 0.63(-4.7) & 0.73(-6.7) & 0.11(-0.3) & 0.17(-0.5) & -0.35(-1.5) \\
         & 25 & -0.14(-0.3) & -0.01(-0.0) & -0.12(-0.3) & 0.72(-5.3) & 0.71(-5.2) & -0.13(-0.3) & -0.17(-0.5) & -0.10(-0.2) \\
        \hline
    \end{tabular}
\end{table}

According to the findings, VTEC and the geomagnetic parameters (Kp, Ap, and Dst indices) consistently show a moderate to weak correlation (inverse for Dst), especially during the ascending and descending periods of solar cycles 23 to 25. During these transitional periods, HSSs and CME events frequently generate increased geomagnetic activity \citep{Astafyeva2015}. The Kp and Ap indices also tend to peak during the descending and minimum phases of the solar cycle due to the increased occurrence of HSSs from coronal holes, which can intensify geomagnetic disturbances and subsequently lead to ionospheric variability \citep{Tsurutani2006b, Verbanac2011}. Storm-time ionospheric disturbances that lower electron density, particularly at mid-latitudes, cause a negative correlation in the Dst index, and a positive correlation for Kp and Ap indices. These low correlations imply that although geomagnetic storms have an impact on VTEC, diurnal variations and solar EUV remain as the main determinants. As ionospheric variability is enhanced by more frequent geomagnetic disturbances, the correlations during ascending and descending phases may be significantly greater. Maximum and minimum phases, on the other hand, exhibit weaker relationships with low statistical significance as ionospheric behaviour during these periods is dominated by non-solar drivers and continuous strong solar forcing respectively.

The geomagnetic parameters and the solar wind flow pressure in solar cycle 24 also exhibit comparatively greater connections with VTEC during the descending phase than in cycle 23. This difference can be attributed to the distinct dominant drivers in each cycle. The descending phase of cycle 24 was characterized by more regular, recurrent geomagnetic forcing from coronal hole high-speed streams (HSS) which create CIRs, thereby raising the dynamic pressure and which then causes frequent, moderate geomagnetic disruptions \citep{Heinemann2020, Heinemann2024, Luhmann2022}. This persistent and moderate pattern facilitated a clearer correlation with global VTEC. Especially at mid-latitudes over China and its neighbouring countries, these persistent disturbances most likely increased ionospheric variability in a quantifiable manner \citep{Gupta2021}. In contrast, the descending phase of solar cycle 23 was dominated by extreme, sporadic events like the 2003 Halloween Storm \citep{Tsurutani2006a}. Such intense, isolated CME-driven storms cause massive, non-linear ionospheric perturbations that disrupt clear correlation patterns in long-term averages, explaining the weaker observed connection.

The poor connection reported between VTEC and both the major solar and geomagnetic indicators during solar cycle 25 is most likely owing to the very low solar activity, particularly during the ascending and minimum phases. Compared to solar cycles 23 and 24, cycle 25 has much fewer sunspots and lower $F_{10.7}$ solar radio flux, implying lower EUV emissions required for ionospheric ionisation \citep{Nandy2021}. As indicated by the p-values, many of the correlations in cycle 25 do not reach statistical significance i.e. $p\geq0.05$ in contrast to the significant relationships found in the stronger cycles 23 and 24. This includes the lagged correlations (see Section~\ref{sec:results:lag}), suggesting that even after accounting for optimal response delays, the relationship remain statistically weak or less than significant. Particularly during the maximum phase of cycle 25, the correlations for key solar proxies i.e. R sunspot number and $F_{10.7}$ struck out on reaching statistical significance, highlighting a stark departure from the robust, phase-dependent links established in earlier cycles.

Halo coronal mass ejections (HCMEs) and related solar wind characteristics during cycle 25 are quantitatively similar to those recorded in the previous solar cycles, but more closely related to the results obtained during the weaker solar cycle 24. This shows that while cycle 25 may be slightly more active than cycle 24 in some respects, its overall heliospheric and space weather behaviour is not comparable to that of the far more active cycle 23. As a result, cycle 25 is classified as a moderately weak cycle, with HCME expansion and solar wind dynamics similar to those of cycle 24, rather than indicating a return to the more intense solar conditions in previous cycle(s) \citep{Gopalswamy2023}. Solar proxies such as He~II and Mg~II, which previously showed a high link with VTEC \citep{Vaishnav2019}, showed less fluctuation, reducing their ionospheric influence. In addition, cycle 25 has seen less violent geomagnetic storms and reduced solar wind dynamics, resulting in limited ionosphere perturbations \citep{Hajra2022}. Under such calm solar-geomagnetic conditions, lower atmospheric dynamics characterised by thermospheric winds and tides, are expected to dominate ionospheric behaviour, making clear connections between solar drivers and VTEC difficult to discern and frequently not statistically robust.

\subsection{Time Lag Characteristics}\label{sec:results:lag}

\begin{figure}
    \centering
    \begin{tabular}{cc}
        \includegraphics[width=0.37\textwidth]{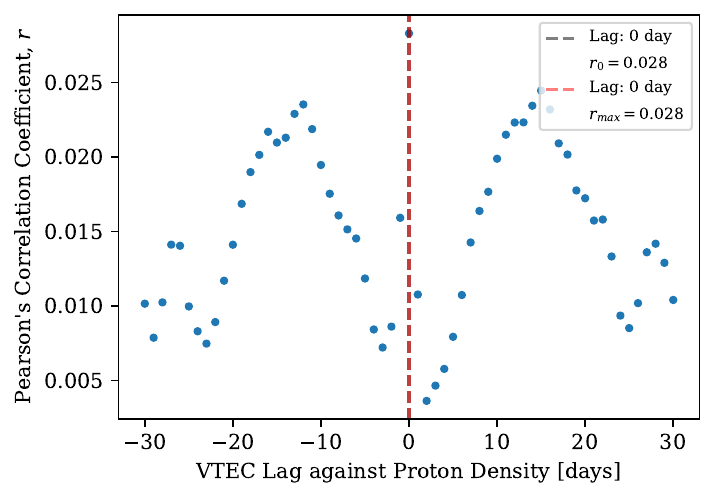} & \includegraphics[width=0.37\textwidth]{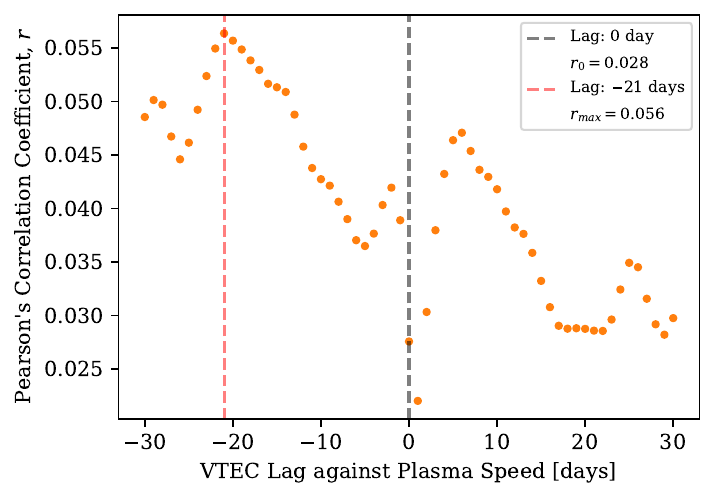} \\
        \includegraphics[width=0.37\textwidth]{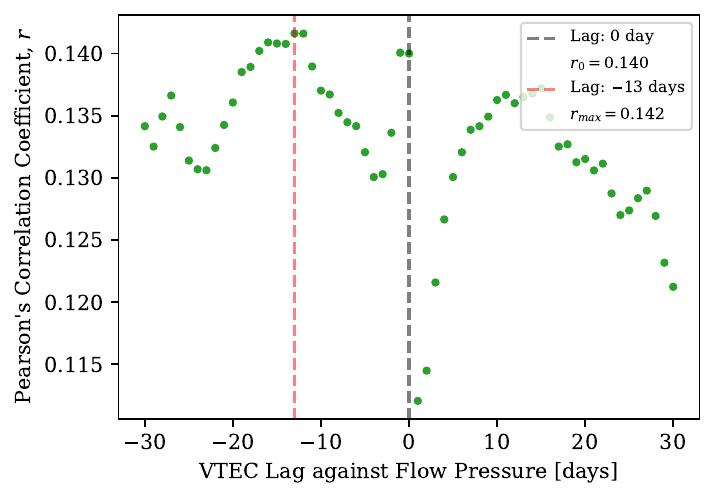} & \includegraphics[width=0.37\textwidth]{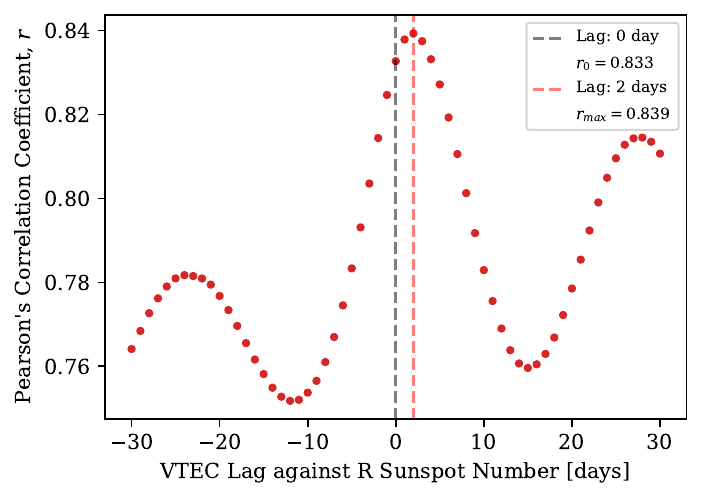} \\
        \includegraphics[width=0.37\textwidth]{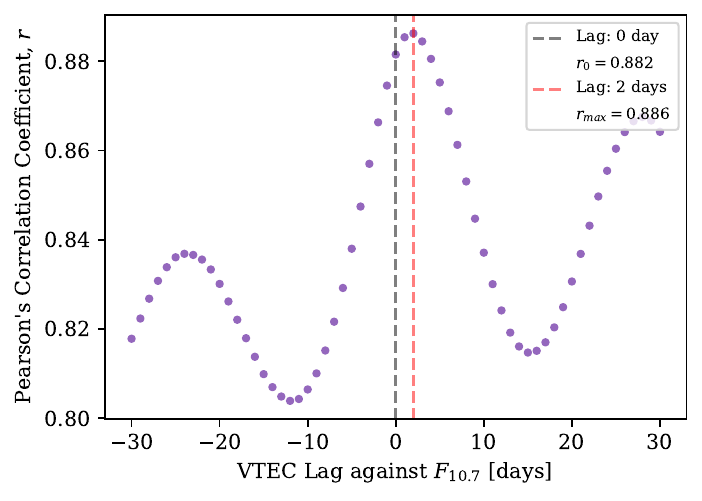} & \includegraphics[width=0.37\textwidth]{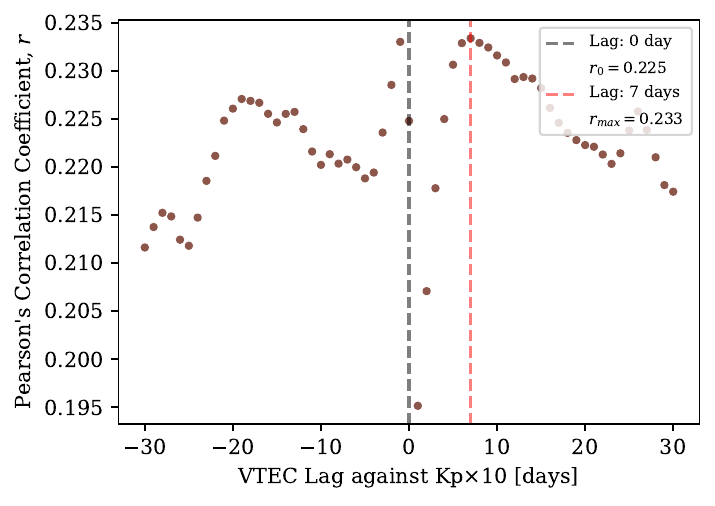} \\
        \includegraphics[width=0.37\textwidth]{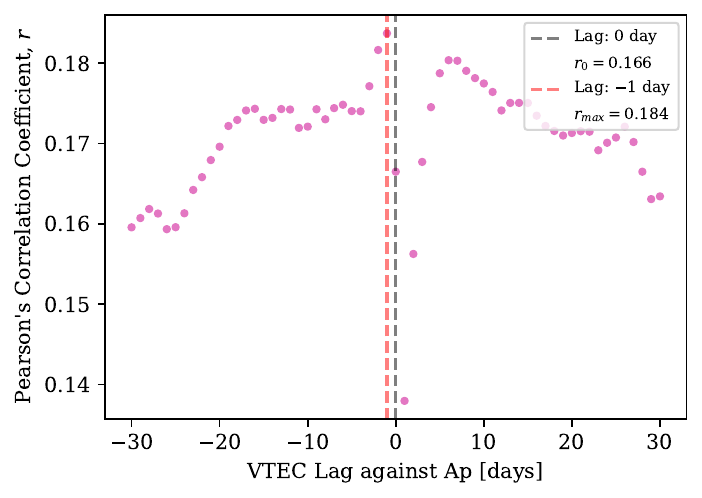} & \includegraphics[width=0.37\textwidth]{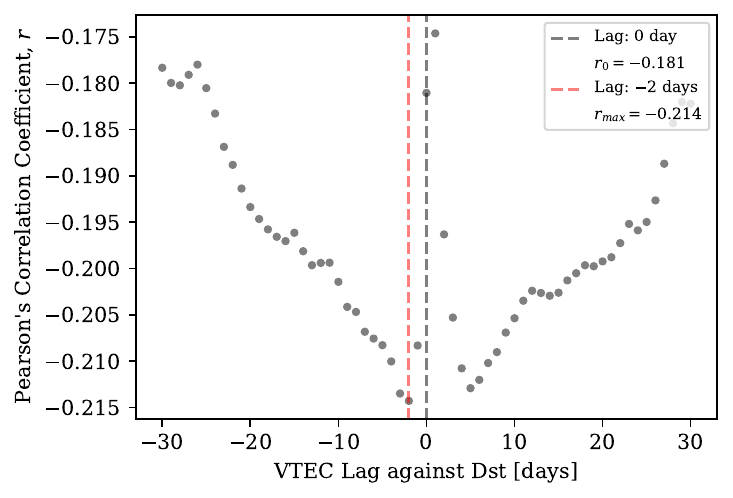}
    \end{tabular}
    \caption{Dependence of the correlation of solar and geomagnetic parameters with VTEC on the VTEC time lag in days. The red dashed vertical line shows the time lag at which the maximum (absolute) correlation coefficient occurs. In cases where the correlations are uniformly weak (e.g. proton density), this peak should not be interpreted as evidence of a meaningful lagged relationship.}
    \label{fig:time_lag}
\end{figure}

Figure \ref{fig:time_lag} shows the dependence of the correlation between the solar and geomagnetic parameter, and the VTEC values, against VTEC lag time in days, while the Table \ref{tab:time_lag-phases} show the VTEC lag time with the maximum absolute correlation for different solar activity phases. Similar to the previous results, the solar activity parameters i.e. R sunspot number and $F_{10.7}$ show strong correlations with daily VTEC values, with maximum correlation occurring at a 2-day lag in ionospheric response. For the solar parameters, there are two primary physical phenomena attributed to the time lag; the behaviour of oxygen in the thermosphere-ionosphere, and the ionisation-recombination balance in the ionosphere. 

\begin{table}
    \centering
    \def\arraystretch{1.2}
    \caption{Time lag for each solar and geomagnetic parameters with VTEC with maximum correlation, for each solar activity phase. The time lag values shown are in days, and the values in brackets are their respective maximum (absolute) correlation coefficient followed by the base-10 logarithm of their respective p-values. For correlations with $p<10^{-250}$, the values are not displayed}.
    \label{tab:time_lag-phases}
    \scriptsize
    \begin{tabular}{cccccccccc}
        \hline
        Phase & Cycle & Proton Density & Plasma Speed & Flow Pressure & R Sunspot No. & $F_{10.7}$ & Kp Index & Ap Index & Dst Index \\
        \hline
        All data & 23 - 25 & 0 (0.03,-2.3) & -21 (0.06,-7.7) & -13 (0.14,-44.4) & 2 (0.84) & 2 (0.89) & 7 (0.23,-121.6) & -1 (0.18,-75.0) & -2 (-0.21,-102.2) \\\hline
        \multirow{3}{*}{Ascending} & 23 & 2 (-0.18,-7.3) & 30 (0.22,-11.1) & 1 (-0.07,-1.6) & 2 (0.65,-112.3) & 2 (0.73,-152.8) & 10 (0.16,-6.2) & 9 (0.10,-2.6) & 5 (-0.09,-2.4) \\
        & 24 & -14 (0.08,-2.0) & -9 (-0.17,-7.1) & 2 (-0.04,-0.6) & 2 (0.75,-172.7) & 1 (0.80,-213.5) & 26 (0.05,-1.0) & 23 (0.10,-2.5) & 22 (-0.17,-6.6) \\
        & 25 & -25 (-0.14,-4.5) & 5 (0.22,-10.1) & 11 (0.08,-1.7) & 2 (0.71,-134.8) & 2 (0.79,-188.3) & 27 (0.27,-15.3) & 8 (0.22,-10.7) & 24 (-0.20,-8.8) \\\hline
        \multirow{3}{*}{Maximum} & 23 & 0 (0.15,-6.3) & 28 (-0.37,-36.4) & 30 (-0.08,-2.1) & 2 (0.49,-68.7) & 2 (0.70,-162.2) & 1 (-0.24,-15.9) & 1 (-0.14,-5.4) & -4 (-0.18,-8.9) \\
        & 24 & -14 (0.05,-1.0) & -24 (0.07,-1.4) & -16 (0.08,-2.1) & 2 (0.49,-61.2) & 2 (0.62,-108.8) & 6 (0.05,-0.9) & 1 (-0.08,-1.8) & 5 (-0.11,-3.4) \\
        & 25 & 8 (0.09,-1.5) & -26 (0.18,-4.6) & -17 (0.11,-2.0) & 30 (0.29,-11.0) & 29 (0.48,-31.0) & 22 (0.21,-5.7) & 23 (0.14,-2.7) & -18 (-0.17,-4.3) \\\hline
        \multirow{2}{*}{Descending} & 23 & 24 (-0.16,-8.3) & -14 (0.26,-20.0) & -14 (0.13,-5.1) & 1 (0.65,-152.6) & 1 (0.73,-213.7) & -13 (0.31,-28.6) & -1 (0.25,-18.5) & -2 (-0.16,-8.0) \\
        & 24 & 0 (0.11,-3.5) & 26 (0.06,-1.4) & 0 (0.21,-10.8) & 2 (0.68,-143.4) & 1 (0.76,-201.1) & 13 (0.23,-13.3) & 14 (0.21,-11.0) & 13 (-0.34,-29.6) \\\hline
        \multirow{2}{*}{Minimum} & 24 & -2 (-0.10,-3.3) & -28 (0.15,-6.6) & 0 (0.14,-6.0) & 2 (0.51,-75.4) & 2 (0.67,-147.5) & 0 (0.19,-10.6) & 0 (0.22,-13.1) & -1 (-0.29,-23.3) \\
        & 25 & -2 (-0.12,-3.7) & 27 (0.15,-5.2) & -3 (-0.08,-2.0) & 2 (0.58,-86.1) & 1 (0.68,-126.4) & 0 (0.12,-3.8) & 27 (0.13,-4.0) & 27 (-0.22,-10.8) \\
        \hline
    \end{tabular}
\end{table}

Atomic oxygen density follows changes in solar radiation, particularly EUV, due to the photodissociation of molecular oxygen at low altitudes. The atomic oxygen, with an average lifetime of several days, then photoionises to produce $\mathrm{O}^+$ ions and electrons (which contribute to VTEC), and this process is dependent on both the solar radiation intensity and atomic oxygen density \citep{Jakowski1991, Jakowski2002}. Furthermore, large-scale thermospheric circulation patterns; such as the disturbance dynamo and equatorward winds triggered by geomagnetic activity would redistribute this atomic oxygen over timescales of one to several days, modulating the net ionization balance in the F-region \citep{FullerRowell1997}. The density dependence induces a varying delay in the ionospheric reaction, and model calculations have shown a time delay of two days between the oscillation of atomic oxygen and initiating solar radiation periodicity \citep{Jakowski1991}. 

The ionospheric delay is also described as an inertial property of the ionosphere i.e. \textit{sluggishness} which is determined by the balance between ionisation and recombination processes \citep{OHare2025}. Photoionisation of atomic oxygen by solar EUV radiation in the F2 region occurs within timescales of less than an hour \citep{deAdler1997}. In contrast, the loss of ionisation is determined by charge exchange of atomic oxygen and molecular nitrogen with $\mathrm{O}^+$ and dissociative recombination \citep{Jakowski1991}, and this process depends on various factors, each with different delay rates \citep[atomic oxygen at 0.3 - 0.6 days, molecular nitrogen at 1.4 - 1.8 days, and their ratio/difference at 3.4 - 5 days;][]{Ren2018}. Ultimately, changes in solar EUV dominate the delay process and result in a shorter overall ionospheric response time. 

Other potential processes that may affect the delay include geomagnetic activities \citep{Kutiev2013}, electrodynamic processes \citep{Vaishnav2019}, atmospheric eddy diffusion \citep{Vaishnav2022a}, seasonal variation \citep{Ren2018}, and lower atmospheric forcing waves \citep{Vaishnav2022a}. Additionally, changes in neutral composition, particularly the O/N$_2$ ratio, play a critical role. During active periods, thermospheric heating and upwelling can decrease this ratio at mid-latitudes over $\sim$1 - 2 days via circulation, enhancing molecular recombination and electron loss \citep{Borries2024}. This adjustment operates on a timescale congruent with the observed lag. All these effects to the ionospheric response are seen in both modelling \citep{Vaishnav2022a} and various physical observations \citep{Kutiev2013, Ren2018, Feng2023}, all with a resulting time lag between the ionosphere and solar activity of approximately 2 days. It is also important to note that there exists a three-peak shape in the correlation to time lag relations (for strong correlators), with two symmetric peaks at approximately 27 days away from the central 2-day peak, and this is consistent with the solar rotation quasi-period which causes the recurrent influence of active solar regions to spread inhomogeneously on the solar surface. 

The R sunspot number and $F_{10.7}$ have the greatest impact during ascending and descending parts of the solar cycle, with a consistent two-day lag, similar to the overall results previously mentioned. This follows the general consensus that EUV plays a major role in increasing the electron content of the ionospheric layer during times of solar activity change \citep{Feng2023}, and highlights how important consistent EUV input is in determining ionospheric behaviour outside of storms. Even when time-lag is applied, correlations between VTEC and other space weather drivers or geomagnetic indicators remain weak. Notably, for certain phases, particularly during solar minimum and for geomagnetic parameters, these time-lag correlations are not statistically significant, indicating that the ionospheric response is not consistently defined under all conditions.

Interestingly, the correlation in the geomagnetic parameters using the daily data also weakens generally, indicating that the VTEC-geomagnetic relationship strengthens over longer averaging timescales i.e. daily rather than monthly in this case. This phenomenon has been observed in various literature \citep[e.g.][]{MoraesSantos2024, Allen2023, Zhai2023, Tyler2024}, even when considering alternative geomagnetic drivers and focusing on specific geomagnetic events \citep{Carmo2024}. This suggests that over short intervals, the ionosphere-thermosphere system is governed by transient, stochastic drivers \citep[e.g. solar wind fluctuations, substorms, gravity waves, tides;][]{Laundal2020}. Conversely, more persistent processes—such as disturbance dynamo and prompt-penetration magnetic fields, meridional neutral winds, and thermospheric compositional changes \citep{Imtiaz2020, Mukhtarov2025, Pal2025}, along with latitudinally influenced seasonal modulation, serve to smooth out daily variability and strengthen the daily geomagnetic-VTEC correlations.

\section{Conclusion}\label{sec:conc}

This study presents a comprehensive analysis of ionospheric total electron content (VTEC) variations in response to solar and geomagnetic influences across solar cycles 23 to 25. Using global CASG VTEC data and NASA OMNI solar-geophysical parameters, we quantified correlation strengths and time-lag dynamics across distinct solar activity phases. Our findings confirm that solar proxies, particularly $F_{10.7}$ flux and R sunspot number, exhibit the strongest and most consistent, statistically significant correlations with VTEC, peaking during ascending and descending phases where solar EUV forcing is more stable. Conversely, geomagnetic parameters such as Kp, Ap, and Dst indices demonstrate weaker and less consistent associations with VTEC with many correlations not reaching statistical significance, modulated primarily by transient events and regional effects.

Phase-resolved analyses, supported by statistical significance testing, revealed that the ionosphere is most responsive to solar drivers during transitional phases of the solar cycle, while responses during maxima and minima were either disrupted by geomagnetic storms or dominated by non-solar influences. Furthermore, we established a $\sim2$-day time lag in VTEC response to solar activity, likely governed by atomic oxygen dynamics and recombination timescales. While direct global correlations between solar wind parameters and VTEC are weak considering that solar wind energy is converted into geomagnetic activity that influences the ionosphere in complex, region-specific ways, localized enhancements that occurs particularly during the descending phase of solar cycle 24, highlight the importance of regional processes and stream interaction events. We also find that VTEC-geomagnetic correlations strengthen at monthly timescales, indicating that persistent, large-scale drivers dominate over short-term, stochastic variability.

Overall, our findings underscore the primacy of solar radiation in modulating ionospheric electron content, while also emphasizing the complexity introduced by geomagnetic and solar wind interactions. The variability in correlation across solar phases and the existence of consistent time delays suggest that predictive ionospheric models must incorporate both solar cycle phase dependence and temporal lag effects. Future studies should pursue higher-resolution regional analyses and multi-instrument comparisons to unravel the spatial heterogeneity in ionospheric responses.

\section*{Acknowledgments}
We acknowledge the support provided by Malaysia’s Ministry of Science, Technology and Innovation (MOSTI) under the Technology Development Fund (TDF07221598). The CASG VTEC data were obtained from \url{ftp://ftp.gipp.org.cn}. The OMNI data were obtained from the GSFC/SPDF OMNIWeb interface at \url{https://omniweb.gsfc.nasa.gov}. The data analysis was conducted using Python, primarily utilizing the \texttt{astropy} package, described on \url{http://www.astropy.org/}.

\bibliographystyle{jasr-model5-names}
\biboptions{authoryear}
\bibliography{ref}

\end{document}